    \documentstyle[nato]{crckapb}

\begin{opening}
\title{THE CHEMICAL EVOLUTION OF THE MILKY WAY }

\author{Monica Tosi}
\institute{Osservatorio Astronomico di Bologna\\
           Via Ranzani 1, I-40127 Bologna, Italy}

\end{opening}

\runningtitle{GALACTIC CHEMICAL EVOLUTION}

\begin{document}

\section{Introduction}

The field of chemical evolution modeling of the Galaxy is experiencing in
the last years a phase of high activity and important achievements.
There are, however, several open questions which still need to be answered.
In this review I will try to summarize what have been the most important
achievements and what are some of the most urgent questions to be answered.

The reason for the recent increase of activity and success of chemical 
evolution models is probably two-folding. First of all, on the observational
side, the last decade has witnessed a tremendous improvement in the 
quality and in the amount of data on the major Galactic features, like the 
chemical abundances and abundance ratios in stellar and gaseous objects of 
various types, the density distributions of gas and stars in different 
Galactic regions, etc.: Fundamental data which provide stringent constraints 
on evolution models.  In addition, also on the theoretical side there has
been a recent blooming of new studies, with several new groups working
on stellar nucleosynthesis to derive reasonable yields for stars of all
mass and of several initial metallicities, and taking into account as much 
as possible the large uncertainties affecting the latest evolutionary 
phases. If we consider that for almost two decades the only usable set of
yields for low and intermediate mass stars was that provided by Renzini \&
Voli (1981), while now we can choose among those by Forestini \& Charbonnel
(1997), van den Hoek \& Groenewegen (1997), Boothroyd \& Sackman (1998)
and Marigo (1998 and this volume), all published in the last two 
years, it is apparent
that we have entered an era of great interest in stellar nucleosynthesis 
studies. 

These circumstances have favoured the appearance in the literature of an
increasing number of good chemical evolution models computed by an increasing
number of people. Nowadays there are several models able to satisfactorily
reproduce all the major observational constraints, not only in the solar
neighbourhood but also in the whole Galaxy. Only in the last few months one 
could count at least four different groups who have presented models in fairly
good agreement with the data: Boissier \& Prantzos (1999, hereinafter BP), 
Chang et al. (1999), Chiappini et al. (1999, CMP) and Portinari \& Chiosi 
(1999, PC).

\section {Major Results} 

Before analysing the various results, it is important to recall that standard
chemical evolution models follow the large-scale, long-term phenomena and
can therefore reproduce only the average trends, not the cloud-to-cloud,
star-to-star fluctuations. To put it in Steve Shore's words: {\it They are 
a way to study the climate, not the weather, in galaxies.} This can be 
considered a limitation of the models, but is the obvious price to pay to avoid 
introducing too many free parameters that would make it much more difficult
to infer the overall evolutionary scenario with sufficient reliability.
As well known, we have not yet been able to find a unique scenario for the
most probable evolution of the Milky Way (see e.g. Tosi 1988a), but we are
converging toward a fairly limited range of possibilities for the involved
parameters (initial mass function, IMF, star formation rate, SFR, gas flows
in and out of the Galaxy).

Thanks to the improvements both on the observational and on the theoretical
sides, good chemical evolution models of the Milky Way nowadays can reproduce
the following list of observed features:

\begin{itemize}
\item Current distribution with Galactocentric distance of the SFR (e.g. as
 compiled by Lacey \& Fall 1985);
\item current distribution with Galactocentric distance of the gas density
 (see e.g. Tosi, 1996, BP and references therein);
\item current distribution with Galactocentric distance of the star density
 (see e.g. Tosi, 1996, BP and references therein);
\item current distribution with Galactocentric distance of element abundances
 as derived from HII regions and from B-stars (e.g. Shaver et al. 1983,
 Smartt \& Rollerston 1997);
\item distribution with Galactocentric distance of element abundances at
 slightly older epochs, as derived from PNe II (e.g. Pasquali \& Perinotto
 1993, Maciel \& Chiappini 1994, Maciel \& K\"oppen 1994);
\item age-metallicity relation not only in the solar neighbourhood but also
 at other distances from the center (e.g. Edvardsson et al. 1993);
\item metallicity distribution of G-dwarfs in the solar neighbourhood (e.g.
 Rocha-Pinto \& Maciel 1996);
\item local Present-Day-Mass-Function (PDMF, e.g. Scalo 1986, Kroupa et al.
 1993);
\item relative abundance ratios (e.g. [O/Fe] vs [Fe/H]) in disk and halo
 stars (e.g. Barbuy 1988, Edvardsson et al. 1993, Israelian et al. 
 this volume).
\end{itemize}

As mentioned above, the most recent examples of how good models can fit 
the above list of observed Galactic features are given by BP, 
Chang et al. (1999), CMP and PC (see also in this book the contributions 
by Chiappini, by Portinari and by Prantzos).

If one bears in mind that the free parameters involved in the computation
of standard chemical evolution models are essentially the IMF, the law 
for the SFR, and those for gas flows in and out of the Galaxy, it is clear
that the number of observational constraints is finally sufficient to put
significant limits on the parameters. In fact, if we compare 
the results of all the models in better agreement with the largest set 
of empirical data, we see that they roughly agree on the selection of
the values for the major parameters. The conclusions that can be drawn from
such comparison are:

\noindent
$\bullet$ {\bf IMF}: after several sophisticated attempts (e.g. CMP)
 to test if a variable IMF could better fit the data, it is
 found, instead, that a roughly constant IMF is most likely, even if the exact
 slopes and mass ends are still subject of debate.

\noindent
$\bullet$ {\bf SFR}: it cannot be simply and linearly dependent only on
 the gas density; a dependence on the Galactocentric distance
 is necessary, either implicit (e.g. through the total mass density as
 in Tosi 1988a or in Matteucci \& Fran\c cois 1989) or explicit (e.g. as
 in BP). We don't know however what is its actual 
 behaviour (see e.g. Portinari, this volume) or even if it should be 
 considered as fairly continuous or significantly intermittent as recently 
 suggested by  Rocha-Pinto et al. (1999).

\noindent
$\bullet$ {\bf gas flows}: all the models in better agreement with the data
invoke no or negligible galactic winds and a substantial amount of infall of 
metal poor gas (not necessarily primordial, e.g. Tosi 1988b, Matteucci \& 
Fran\c cois 1989) and there are increasing observational evidences on this 
phenomenon (see also Burton, this volume).
We have no empirical information, however, on the spatial and temporal
distribution of the accretion process: uniform or not ? 
continuous or occurring in one, two or several episodes ? (e.g. Beers 
\& Sommer-Larsen 1995, Chiappini et al. 1997, Chang et al. 1999).

\section {Open Questions}

It is apparent from the summary presented above that,
in spite of the wealth of good data and models described in the previous
sections, the scenario of the Milky Way evolution is not completely
clear. There are still several issues we don't understand, including
some of conspicuous importance. Among these, I consider of special interest
the evolution of the abundance gradients and that of CNO isotopes.

\subsection {Abundance Gradients}

Thanks to the recent results by Smartt \& Rollerstone (1997) we finally know
that young objects (HII regions and B-stars) all show the same metallicity
distribution with Galactocentric distance and a fairly steep negative gradient. 
All the models in better agreement with the Galaxy constraints are able to
reproduce this distribution (see Tosi 1996, Chiappini, Portinari and
Prantzos in this volume).

\begin{figure}
\vspace{9truecm}
\includegraphics{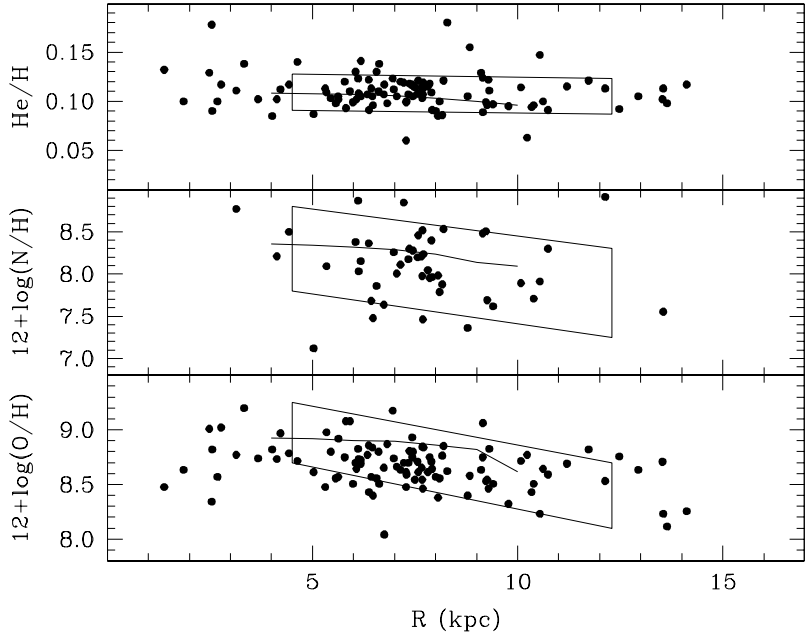}
\caption
{Radial distribution of the He, N and O abundances as derived from observations
of PNe of type II (see text for references) and from the predictions of 
Tosi's (1988a) model 1 for the Galaxy medium 2 Gyr ago. }
\label{pne} \end{figure}

Slightly older objects, such as PNe of type II whose progenitors on average 
are 2 Gyr old, show similar abundances and possibly flatter gradients 
(e.g. Maciel \& K\"oppen 1994).
Good models of Galaxy evolution reproduce well not only the present abundance
distribution, but also
the distributions derived from PNeII observations. For instance,
Fig.\ref{pne} shows the predictions of the best of models of type 1 in
Tosi's (1988a) set for the He,
N and O abundance distributions with Galactic radius 2 Gyr ago. The adopted
stellar yields are Marigo's (1998 and this volume) for low and 
intermediate mass stars and Limongi et al. (this volume) 
for massive stars. The data points correspond to the 
PNeII measures by Pasquali \& Perinotto (1993) and the open boxes sketch 
the distribution of the values derived by Maciel \& Chiappini (1994) and 
Maciel \& K\"oppen (1994). The data sets are in perfect agreement with 
each-other and the model predictions fit well their average distributions.

\begin{figure}
\vspace{5.6truecm}
\includegraphics{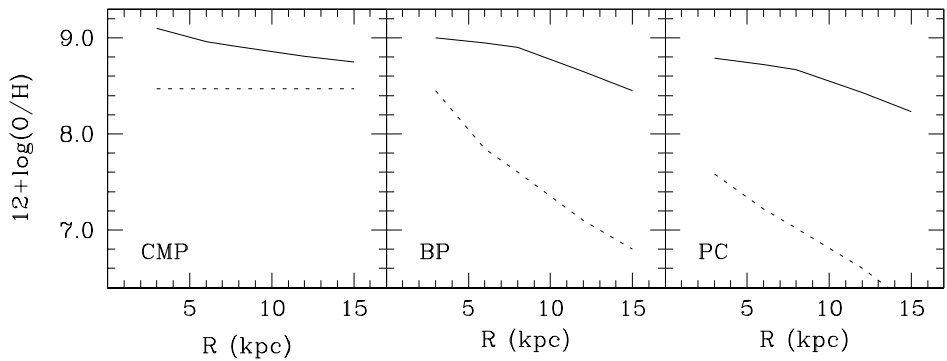}
\caption
{Oxygen gradients predicted by the models from Chiappini et al. (1999, left), 
Boissier \& Prantzos (1999, center), and Portinari \& Chiosi (1999, right).
The solid curves refer to the present epoch and the dotted ones to 1 Gyr after
the disk formation.}
\label{grad} \end{figure}

When we consider earlier epochs, the predictions from different
models diverge, despite the common assumption that the Galaxy is initially 
formed of primordial gas.  For instance, the three models which are 
presented in this volume by Chiappini, Portinari and Prantzos, and that
are in fairly good agreement with all the observational constraints,
predict the gradient evolutions schematically described in Fig.\ref{grad} 
(see BP, CMP and PC for more details). The initial distribution of oxygen with
galactic radius in the left panel is totally flat, becomes
initially slightly positive, then turns to negative and steepens with time, 
reaching at the present epoch the observed slope of -0.08 dex/kpc; vice 
versa, the gradient at 1 Gyr in the central panel 
is negative and quite steep and then slowly flattens with time, 
particularly in the inner galactic regions, reaching finally the observed
slope at the present time; the same trend occurs in the right panel,
but with different absolute abundances.
If one compares (e.g. Tosi
1996) all the models able to reproduce the observed Galactic features, it is
easy to understand that they present all the possible varieties of gradient
evolution: from slopes initially positive becoming first flat and then 
increasingly negative, to slopes initially flat and then becoming increasingly
negative, to slopes initially negative and then becoming increasingly flat.

The reason for such a variety of gradient evolutions is the strong dependence 
of the radial slope on the radial variations of the ratio between ISM
enrichment from stars (i.e. SFR) and ISM dilution from metal poor gas
(i.e. initial conditions and/or infall of metal poor gas). 
Regions with higher SFR have larger
enrichment, but can remain relatively metal poor if they contain or accrete
large amounts of metal poor gas. It is then sufficient to have 
different initial conditions or different assumptions on the temporal
behaviours of the SFR and of the infall rate to obtain quite different
abundance gradients at the various epochs.

The following few examples of possible scenarios give an idea of the 
sensitivity of the gradient evolution to the boundary conditions:
\begin{itemize}

\item If the efficiency in the chemical enrichment of the inner Galactic
regions at early epochs is low (for instance because the SFR is low and/or
there is a high amount of primordial gas), then the early radial
distribution of the heavy elements is flat. And to reach the observed
present slope it has to become negative and steepen with time.

\item If, instead, the enrichment efficiency in the inner regions at
early epochs is high (for high SFR or low gas mass), then the early
gradient is negative and steep. And to reach the present slope it has to 
flatten with time.

\item If at late epochs the acretion (infall) of metal poor gas is stronger 
in the outer than in the inner regions, then the gradient tends to steepen 
with time because of the increasing dilution for increasing galactocentric
distance.

\item If at late epochs the inner regions exhaust their gas, then the 
metallicity saturates there and the inner gradient becomes increasingly flat
with time.

\end{itemize}

All these scenarios are plausible: how can we understand which are the right 
ones ? If we knew the right history of the abundance gradients we would
also know what is the most likely evolution of the Galactic disk. Unfortunately,
despite their accuracy, the observational data already available on open
clusters and on field stars are not yet sufficient to clearly distinguish
whether the abundance gradients were steeper or flatter at early epochs.
Open clusters are probably the best candidates to provide such information,
thanks to their visibility at large distances and to the relative ease
to derive their age and metallicity, but as described by Bragaglia (this
volume, and references therein) the number of clusters treated homogeneously 
is still too small.

\subsection {Evolution of CNO isotopes}

The CNO isotopes are important because they are stable, diffused and largely
studied, since they provide the seeds for the production of heavier
elements. In particular, the stellar nucleosynthesis of the carbon and oxygen
isotopes is examined in detail in most of the most recent studies. 
Nonetheless, it is not completely clear yet how they should behave during
the Galaxy evolution. The problem was already pointed out twenty years ago
by Penzias (1980), who noticed that the observed decrease of the local
$^{18}$O/$^{17}$O from the solar to the local ISM value and the corresponding
increase of $^{16}$O/$^{18}$O were
difficult to interpret. In fact, chemical evolution models predicted
(Tosi 1982) $^{18}$O/$^{17}$O to remain roughly constant in the last 4.5 Gyr 
and $^{16}$O/$^{18}$O to steadily decrease. 
Those predictions were based on simple arguments on the relative
enrichment of primary and secondary elements produced by stars of different
masses, and have been confirmed by subsequent studies based on 
nucleosynthesis studies of solar metallicity stars (e.g. Prantzos
et al. 1996). 

\begin{figure}
\vspace{13.5truecm}
\includegraphics{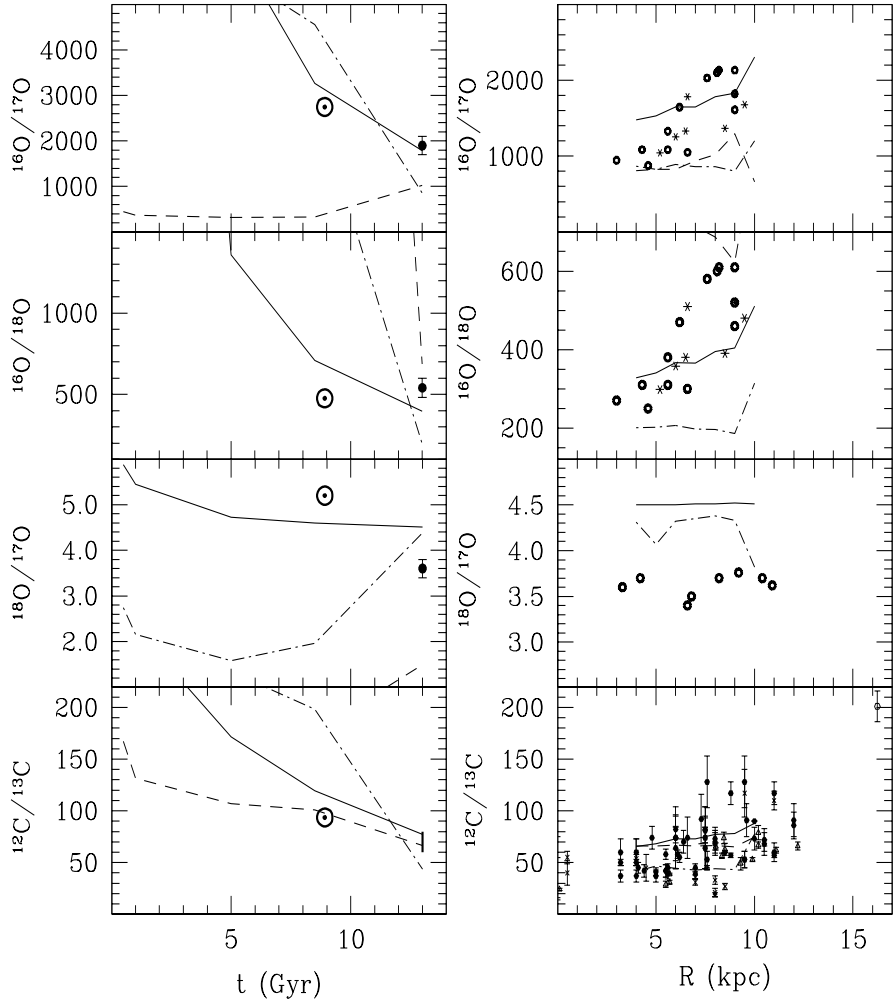}
\caption
{Carbon (bottom panels) and oxygen (three top panels) isotopic ratios.
Left panels: evolution in the solar neighbourhood. Right panels: current 
distributions with Galactocentric distance. 
The solar symbol represents the solar ratio derived from Anders \& Grevesse 
1989; all the other data are from radio observations of molecular clouds 
(see Sandrelli et al. 1998, for references).
All the curves refer to Tosi (1988a) model 1 but assuming different stellar 
yields as described in the text.
}
\label{isotopes} \end{figure}

These results for the carbon and oxygen isotopic ratios are represented
by the solid line in Fig.\ref{isotopes}. The left hand panels show the
time behaviour of the isotopic ratio in the solar neighbourhood as predicted
by  models and as observed in the sun and in the local ISM, which are assumed
to be representative of the average local ratios 4.5 Gyr ago and now,
respectively. 
The right hand panels show the present distribution with Galactocentric 
distance as predicted by the same models and as derived from radio 
observations of molecular clouds. 
The solid line corresponds to the same model presented
in Fig.\ref{pne} (Tosi-1), assuming the yields for solar initial metallicity
computed by Boothroyd \& Sackman (1998), by Forestini \& Charbonnel (1997)
and by Woosley \& Weaver (1995) for low, intermediate and high mass stars,
respectively. Qualitatively similar results were obtained by Prantzos et al.
(1996) adopting the solar yields by Marigo et al. (1996), Renzini \& Voli 
(1981) and Woosley \& Weaver (1995).
It is apparent that while the predictions for $^{12}$C/$^{13}$C and 
$^{16}$O/$^{17}$O are in fair agreement with the data, the time behaviour
of the oxygen isotopic ratios involving $^{18}$O is inconsistent with them.
There have been several speculations on how this impasse could be overcome, 
with suggestions that either the theory or the data or both might be wrong or
misinterpreted (see e.g. Prantzos et al. 1996, Tosi 1996, Wielen \& Wilson 
1998), but no solution has been found yet.

One possibility is that it is not correct to adopt solar yields also for
the earlier epochs, when stars were certainly metal poorer. Now that stellar
yields are available also for lower metallicities, we expect to find an
improvement in the comparison between model predictions and observed ratios.
Unfortunately, this is definitely not the case, as clearly shown by the
dashed and dash-dotted lines in Fig.\ref{isotopes}. The dash-dotted curve
represents the same model as the solid curve, with the same sources for
the yields, but adopting the low metallicity yields at earlier epochs and
the solar ones only when the ISM reaches Z=0.02.  It is apparent that,
rather than improving the agreement with the data, this curve worsens the
fit, both for the local evolution and for the current distribution
with Galactocentric distance. This result is strongly dependent on the
adopted yields and we may hope that different nucleosynthesis studies
would provide more consistent predictions, but so far no set of stellar
yields is able to reproduce all the shown observed distributions. Some of
the available yields do improve the results on one isotopic ratio, but
worsen the results on other ratios, as exemplified by the dashed
lines, showing the predictions of the same model when Marigo's (this volume) 
metallicity dependent yields are adopted for low and intermediate mass
stars and Limongi's et al. (this volume) for massive stars: 
the data on the carbon isotopic ratio are now well reproduced, but the 
predicted oxygen ratios are definitely inconsistent with the data.

\vskip 0.5truecm

I will then conclude this short description of the state of the art in
Galactic chemical evolution models by emphasizing that, despite the great
work that has been done by observers and theoreticians to improve the
number and the quality of the observational and theoretical constraints,
further efforts on both sides are needed to shed light on several unclear
issues. In particular, it would be important to derive accurate chemical 
abundances in stars and clusters of different ages and Galactic locations
and to study in better detail the stellar nucleosynthesis in stars of all
masses and initial metallicities.

\vskip 0.5truecm

I warmly thank S.Chieffi, M.Limongi, P.Marigo and O.Straniero for providing
their yields in advance of publication and S.Sandrelli for help. Conversations
with them and with C.Chiappini, F.Matteucci and N.Prantzos were very
fruitful.

{}

\end{document}